\documentclass[conference, compsoc]{IEEEtran}
\usepackage[backend=bibtex,bibstyle=ieee,citestyle=numeric-comp]{biblatex}
\AtBeginBibliography{\small}
\bibliography{bibliography}
\IEEEoverridecommandlockouts
\usepackage{amsmath,amssymb,amsfonts}
\usepackage{algorithmic}
\usepackage{graphicx}
\usepackage{textcomp}
\usepackage{xcolor}
\usepackage{mathtools}
\usepackage{dsfont}
\usepackage{tabularx}
\usepackage{multirow}
\usepackage[normalem]{ulem}
\usepackage{todonotes}
\setuptodonotes{inline}
\newcolumntype{Y}{>{\centering\arraybackslash}X}
\usepackage{caption} 
\newcommand{\linebreakand}{%
  \end{@IEEEauthorhalign}
  \hfill\mbox{}\par
  \mbox{}\hfill\begin{@IEEEauthorhalign}
}

\newcommand{\HG}[1]{\textcolor{blue}{#1}}

\title{Risk stratification of malignant melanoma using neural networks}

\author{
\IEEEauthorblockN{Julian Burghoff}
\IEEEauthorblockA{\textit{Dept Mathematics and IZMD} \\
\textit{University of Wuppertal}\\
Wuppertal, Germany \\
burghoff@math.uni-wuppertal.de}

\and

\IEEEauthorblockN{Leonhard Ackermann}
\IEEEauthorblockA{\textit{BioCap GmbH}\\
Frankfurt, Germany \\
l.ackermann@biocap.de}

\and

\IEEEauthorblockN{Younes Salahdine}
\IEEEauthorblockA{\textit{NeraCare GmbH}\\
Frankfurt, Germany \\
younes.salahdine@neracare.com}

\and

\IEEEauthorblockN{Veronika Bram}
\IEEEauthorblockA{\textit{NeraCare GmbH}\\
Frankfurt, Germany \\
Veronika.bram@neracare.com}

\and

\IEEEauthorblockN{Katharina Wunderlich}
\IEEEauthorblockA{\textit{CentroDerm GmbH}\\
Wuppertal, Germany \\
Katharina.wunderlich@gmx.net}

\and

\IEEEauthorblockN{Julius Balkenhol}
\IEEEauthorblockA{\textit{CentroDerm GmbH}\\
Wuppertal, Germany \\
j.balkenhol@centroderm.de}

\and

\IEEEauthorblockN{Prof. Dr. Thomas Dirschka}
\IEEEauthorblockA{\textit{CentroDerm GmbH}\\
Wuppertal, Germany \\
t.dirschka@centroderm.de}

\and

\IEEEauthorblockN{Prof. Dr. Hanno Gottschalk}
\IEEEauthorblockA{\textit{Institute of Mathematics} \\
\textit{TU-Berlin},
Berlin, Germany \\
gottschalk@math.tu-berlin.de}
}

\begin{document}

\maketitle

\begin{abstract}
In order to improve the detection and classification of malignant melanoma, this paper describes an image-based method that can achieve AUROC values of up to 0.78 without additional clinical information. Furthermore, the importance of the domain gap between two different image sources is considered, as it is important to create usability independent of hardware components such as the high-resolution scanner used. Since for the application of machine learning methods, alterations of scanner-specific properties such as brightness, contrast or sharpness can have strong (negative) effects on the quality of the prediction methods, two ways to overcome this domain gap are discussed in this paper.
\end{abstract}


\section{Introduction}

Clinical and pathological staging of melanoma patients only relies on tumor size (Breslow thickness), ulceration and lymph node involvement \cite{gershenwald2018melanoma}. However, the patient group with the thinnest melanomas in Tumor Stage T1 and with the most favorable prognosis resulted the most melanoma deaths in absolute numbers \cite{landow2017mortality}. Additionally, patients from T3b onwards can be now offered potent adjuvant therapy \cite{luke2022pembrolizumab,bms2022}. 
To identify patients with small tumors but high risk of relapse or to spare patients in advanced disease but with low mortality or recurrence risk, there is a current need for biomarkers and better prognostication \cite{rizk2020biomarkers}.
The use of digitalized histological images like H\&E scans, that help pathologists to better interpret the information provided by the tissue sample under the microscope, has been widely tried to use for improving diagnosis and prognosis in many tumors \cite{gurcan2009histopathological}. 3D high-resolution volumetric imaging of tissue architecture from large tissue and molecular structures at nanometer resolution are new techniques for improving early cancer detection, personalized risk assessment and potentially identifying the best treatment strategies \cite{liu2019high}. Deep learning has been shown to read out additional information of these stains. These models have shown to deliver independent prognostic information\cite{qaiser2022usability,combalia2022validation}  and could also predict the results of molecular biomarkers \cite{lee2022deep}.

In melanoma, CNN networks were shown to reach a concordance level above 80\% for diagnosis compared to human pathologists\cite{hekler2019pathologist}, could outperform histopathologists in classification\cite{hekler2019deep}, and the result of the sentinel lymph node status \cite{brinker2021deep}.

Prognostication and predicting the risk of tumor recurrence could be shown for combining digital analysis with the detection of tumor-infiltrating lymphocytes by achieving a negative predicting value (NPV) of about 85\% \cite{moore2021automated}.  Another CNN approach for predicting disease specific survival in melanoma resulted in mixed results achieving area under the curve (AUROC) values of 90\% and 88\% but only NPVs of 95\% and 65\%, respectively, in two validation cohorts \cite{kulkarni2020deep}.

Such problems of the domain gap often arises in image recognition \cite{goodfellow2016deep}, i.e. a method that has been optimised on a data set from a certain source, but on data from a different source it provides unsatisfactory results. To address this problem, we first standardise the predictions for each dataset or data source, which already improves the accuracy. In a second approach, we add an additional regularisation term directly to the neural network so that the results are in the same range. This also improves the accuracy of the predictions.

The following parts of this paper are structured as follows: 
First, in section \ref{sec:DataDescription} we give a description of the data in terms of its clinical and technical structure. Section \ref{sec:MaterialsAndMethods} provides an overview of methods that have been used in similar work and how our method is designed and built on these. Then, in section \ref{sec:results_all}, we present and classify the results of our experiments. Finally, section \ref{sec:discussion} discusses the results and considers how future work might build on them.


\section{Description of Data}
\label{sec:DataDescription}
In this paper we consider two main sets of data:
\begin{itemize}
    \item \textit{Dataset A}: This dataset contains 767 images of 176 patients of the American Joint Committee on Cancer (AJCC) set stages IA to IIID from four different locations in Bern (Switzerland), Bochum, Bonn and Kiel, Germany. All images have been recorded with the H\&E colouring and were created by the same scanner Hamatasu NanoZoomer S210, NDP Version 2.4. 

    In order to obtain a data set of the highest possible quality, the data were first manually checked by medical experts and a total of 23 data points were removed from the data set, e.g.\ due to broken slides. This clean-up took place before we split the data into training, validation and test data:

    The patients are assigned to train (104 patients), test (36 patients) and validation (36 patients) what results in 591 train-, 74 validation- and 102 test images. The split was chosen so that the distribution of high/low risk patients in each subset (training, validation, test) corresponds to the distribution of the full dataset A, see also table \ref{tab:clinical_table}, under the constraint that multiple images of one patient remain in the same subset.
    
    \item \textit{Dataset B}: The second dataset contains 242 images of 242 patients with AJCC stages IIA to IIC from the Central Malignant Melanoma Registry (CMMR) in Tübingen, Germany, where the H\&E-colored images have partly different coloring than the images of \textit{Dataset A}, probably caused by another scanner type Hamatasu Nanozoomer 2.0 HAT, NDP Version 2.5 and the scans are mostly disturbed by a marker pen on the slide. To see how our algorithm performs on images with a domain gap \textit{dataset B} is only used as a separate test dataset.
\end{itemize}

\subsection{Clinical Description}
\label{subsec:DataDesc_Clinical}

The images used represent hematoxylin-eosin (HE) stained melanoma sections. HE is a staining technique from histology that is used to better predict the disease prognosis of a melanoma patient, among other things, the mitosis rate can be determined (S3 Leitlinie Melanom).

Overall survival of dataset A (86\%) is similar to dataset B (88\%). However, the MSS time for dataset B with a median of 41 month was considerably lower compared to 70/ 73 months of dataset A. Relapse free survival differed as well with around 78\% and 69\% for dataset A and dataset B, respectively.

\begin{table*}[]
\centering
 \begin{tabular}{|c|c c c|c|} 
 \hline
 Characteristic & & Dataset A & & Dataset B \\
  & Training & Validation & Test & Test \\ 
 \hline
 N(Scans)         & 313 & 36 & 38 & 242 \\
 N(Patients)      & 104 & 36 & 36 & 242 \\
 \hline
 Alive / Censored & 89 (86\%) & 31 (86\%) & 31 (86\%) & 212 (88\%) \\
 Dead / Event     & 15 (14\%) & 5 (14\%) & 5 (14\%) & 30 (12\%) \\
 MSS time (months) Mean & 70.12 & 80.03 & 78.37 & 50 \\
 MSS time (months) Median & 70 & 73 & 73 & 41 \\
 \hline
 Relapse Free Survival Recurrence & 21 (20\%) & 8 (22\%) & 7 (19\%) & 75 (31\%) \\
 Relapse Free Survival Non-recurrence & 83 (80\%) & 28 (78\%) & 29 (81\%) & 167 (69\%) \\ 
 RFS time Mean & 67 & 76.67 & 75 & 42 \\
 RFS time Median & 70 & 68 & 73 & 30 \\
 \hline
 \end{tabular}
 \caption{Important features of our datasets.}
 \label{tab:clinical_table}
\end{table*}


\subsection{Technical Description}
\label{subsec:DataDesc_Technical}
Both datasets contain a total of 1009 images of different sizes (up to a resolution of $158720 \times 115456$ pixels) and various format ratios. Therefore\HG{,} the images have to be pre-processed as neural nets on commercially available hardware are not yet able to handle images of this size at the time of writing. See section \ref{subsec:MaM_Preprocessing} for more information on the pre-processing task. The total size of the dataset is 570,3 GB in \texttt{.ndpi} file format.
 
For every patient $i$\HG{,} we have at least one image $x_i$ and the information $\delta_i=1$, indicating the death of the patient at time $T_i$. If the patient survived the observation period of this study, we set $\delta_i = 0$ and $T_i$ is the time the patient has been observed.

\section{Materials and Methods}
\label{sec:MaterialsAndMethods}

\subsection{Related Work}
\label{subsec:MaM_RelatedWork}
There exists abundant work on using deep neural networks for the interpretation of medical images. In this section, we focus on works with the scope of AI based prediction of survival for patients with a melanoma diagnosis.

In prior research on this topic \cite{kulkarni2020deep} the authors use a two stage pipeline to predict risk on the basis of primary melanoma tumor images. Within this pipeline, they first use a segmentation to classify detect and crop tumor areas in the image. These small but detailed crops of $500\times 500$px  are then fed to a network of convolutional, recurrent  and fully connected layers in order to predict the risk. 
\cite{johannet2021using} describes the approach to use a multivariable classifier that contains, besides clinical data, a score of a deep neural net, in order to predict the immunotherapy response of patients with advanced melanoma. Also here, clinical data is used for the model and a separation of segmentation and response classifier takes place.

The authors of \cite{hekler2019deep} describe an experiment where a trained ResNet50 model outperforms 11 pathologists in classifying labeled histopathological images what shows that neural nets in general have high potential to improve correct melanoma diagnoses.

Similar to our approach, the authors of \cite{li2022application} used a VGG-based neural network architecture to detect cutaneous melanoma, although they use a binary classification for dead/alive patients instead of a survival analysis method. They evaluated their method a dataset provided by The Cancer Imaging Archive of 53 patients with a given survival status. 

In \cite{sheikhzadeh2016automatic} the authors on the one hand locate molecular biomarkers in immunohistochemistry images using convolutional neural networks which can help enabling new cancer screenings. On the other hand they also classify the found biomarkers along their type.

However, there are not only machine learning approaches to melanoma classification based on H\&E scans, but also based on topics such as dermoscopic image data \cite{patil2022machine} where the authors use a quiet small 5-layers CNN to classify the images to the corresponding tumor stage with applying the Adam optimizer on the Similarity Measure for Text Processing as loss function. This work is also based on \cite{jaworek2019melanoma} where the authors predict the melanomas thickness using a pretrained VGG-19 model on $400 \times 400$px preprocessed dermoscopic images.


In \cite{zhu2016deep} the authors also use deep convolutional neural networks in combination with survival analysis in order to find good predictions on pathological images - here in context of lung cancer. They annotated the regions of interest of the images with help of pathologists and sampled small random high resolution crops of these regions to use in the networks whereas we used down-sampled low resolution images in our approach.



In contrast to this, in our work we extract information directly from the original image data using a VGG16-like neural network, so we don't use an additional pre-segmentation, which might be error prone by itself. In this way we also avoid labeling of the regions of interest by humans annotators, which is a time consuming process and requires highly trained annotators.

\subsection{Preprocessing the Data}
\label{subsec:MaM_Preprocessing}

When pre-processed, the images are reduced from their original format by a factor of 64 in dataset A and 128 in dataset B in each dimension. The resulting image is centred in a 2500 by 2000 pixel frame which is filled with white color outside the image. 

Images of patients with multiple images are seen as independent information in the training data.


\subsection{Methods}
\subsubsection{Cox's Proportional Hazards Model}
Survival analysis is about predicting the probability of absence of an event (e.g. the death of a patient) until time $t$ using the parameters $\beta$ of a suitable model \cite{harrell2001regression}. One of the most widely used methods is Cox's Proportional Hazards Model, which predicts the hazard function $h(t|x)$ on the basis of an input vector $x$ , see e.g., \cite{kulkarni2020deep}, \cite{zhu2016deep} or \cite{Poelsterl2020} :
\begin{equation}
    h(t|x)=h_0(t)\cdot \exp(\beta^T x) \Leftrightarrow \log \frac{h(t|x)}{h_0(t)} = \beta^T x
\end{equation}
The baseline hazard function $h_0(t)$ indicates how large the hazard rate would be without the influence of other parameters (like $\beta$) and is therefore only dependent on the time $t$, which means that for our case it is eliminated from the calculation of the loss function and therefore does not need to be calculated \cite{Poelsterl2020}.
To estimate the parameters $\beta$ of the linear model the negative log partial likelihood function can be minimized:
\begin{equation}
    l(\beta) = - \sum\limits_{i=1}^n\delta_i(\beta^T x_i - \log \sum\limits_{j\in R_i}\exp(\beta^T x_j))
\end{equation}
where $\delta$ is the delta-function which determines whether the data is censored or not 
, $n$ the total number of data points  and $R_i = \{j | y_j \geq y_i\}$ is the risk set which describes the data-subset of patients which do not had an event before timestamp $y_i$ of the $i$-th event.  

An adaption of linear models to non-linear models like (deep) neural nets was introduced by \cite{faraggi1995neural} in 1995 and applications can be found for example in \cite{zhu2016deep} or \cite{Poelsterl2020}. The risk function $\beta^T x$ is replaced by the output of the neural net $\hat{h}_\theta(x)$ and we achieve the following loss function for each mini batch $B$:
\begin{equation} \label{eq:loss_function}
    \mathcal{L}(\theta) \coloneqq -\sum\limits_{i\in B} \delta_i \left(\hat{h}_{\theta} (x_i) - \log \sum\limits_{j\in R_i} e^{\hat{h}_{\theta} (x_j)}\right)
\end{equation}

\subsubsection{Neural Net Architecture and Training}

\begin{figure*}
    \centering
    \includegraphics[width=0.93\linewidth]{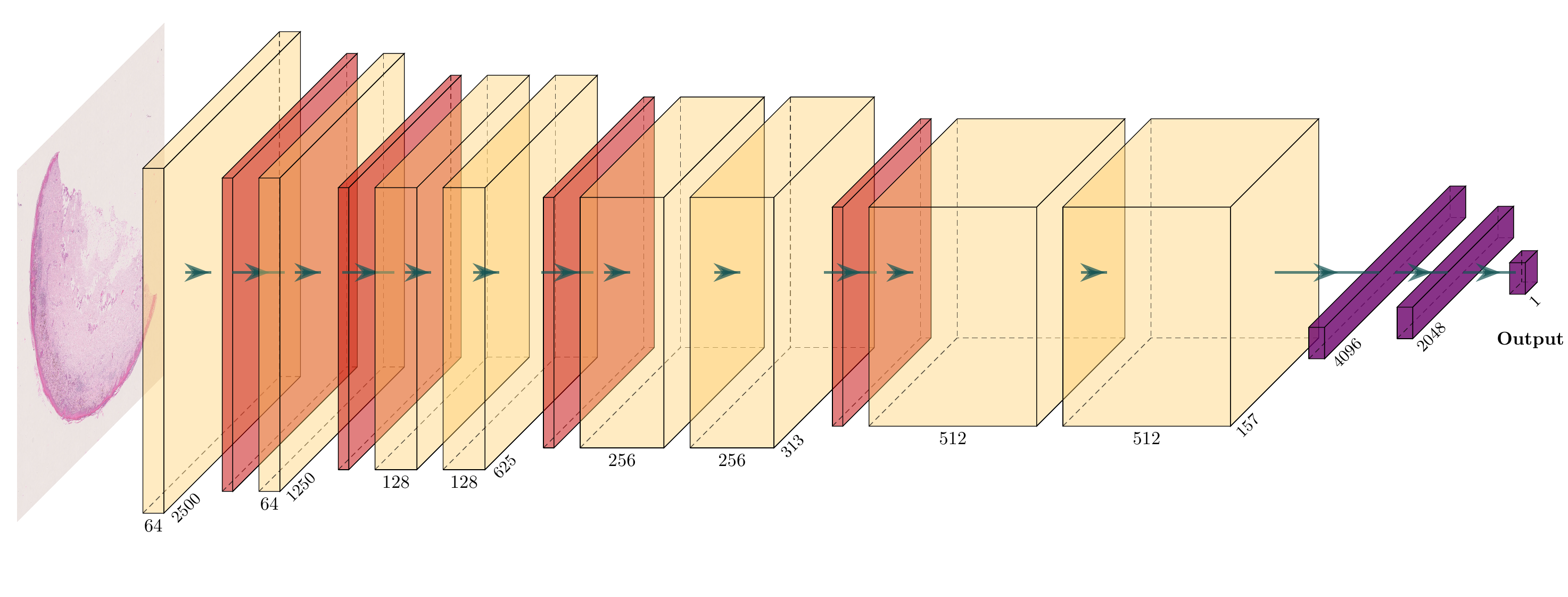}
    \caption{Design of used neural network. Orange are convolutional layers, red are pooling layers and purple indices dense layers.}
    \label{fig:design_neuralnet}
\end{figure*}

Convolutional neural networks (CNN) represent the state of the art in image recognition. As neural network model, we use a modified version of a VGG16 net \cite{simonyan2014deep}  in our experiments. Figure \ref{fig:design_neuralnet} gives an overview of the network architecture. Each convolutional layer has kernel sizes of $3\times 3$ and each pooling layer is a maximum pooling with pooling size $2\times 2$.
    
As mentioned in section \ref{subsec:MaM_RelatedWork}, we use the loss function supplied in \cite{Poelsterl2020} which implements a variant of the Cox's Proportional Hazards Model. In order to achieve a higher generalisability and reducing the domain gap between dataset A and dataset B, we expanded our loss function in \ref{eq:loss_function} by a regularization term:
\begin{multline} \label{eq:lossWithReg}
    \mathcal{L}(\theta) \coloneqq -\sum\limits_{i\in B} \delta_i (\hat{h}_{\theta} (x_i) - \log \sum\limits_{j\in R_i} e^{\hat{h}_{\theta} (x_j)}) \\ + \lambda ( \frac{1}{|B|} (\sum\limits_{j \in B}{\hat{h}}_{\theta} (x_j))^2
\end{multline}
where $\lambda$ is the chosen regularization strength. See sec \ref{sec:results_without_reg} for further information why we use this additional regularization term.


In training, we employ the Adam optimizer over $50$ epochs with a learning rate of $\alpha = 0.001$ and a batch size of $5$. 

For the implementation we used Openslide (version 1.1.2) \cite{goode2013openslide} for importing the images, the GitHub repository of Sebp\cite{Poelsterl2020} and Tensorflow (version 2.6.1) \cite{tensorflow2015-whitepaper} along with Keras (version 2.6.0) \cite{chollet2015keras}.

For the calculations, we used a workstation with a Dual Intel Xeon Gold 6248R 3.0GHz and three Nvidia Quadro RTX 8000 graphic units with 48GB VRAM each, whereas the different GPUs are only used for different trainings.

\subsubsection{Concordance Index}
For validation purposes of our methods we use Harrel's concordance index (CI) \cite{HarrelsCIndexOrig}. 
The CI is defined as the ratio between correctly ordered pairs and all possible rankable pairs \cite{schmid2016CIndex}:
\begin{equation}
    CI = \frac{\#\text{concordant pairs}}{\#\text{comparable pairs}}
\end{equation}
A pair of observations $i,j$ with its survival times fulfill $T_i > T_j$, is concordant if $\hat{h}_{\theta} (x_j) > \hat{h}_{\theta} (x_i)$. 
Also a pair $i,j$ is not comparable if the smaller survival time is censored (i.e. $T_i>T_j \land \Delta_j=0$). Otherwise this pair is comparable. Thus,
\begin{equation}
    CI = \frac{\sum_{i,j}\mathds{1}(T_i > T_j)\cdot\mathds{1}(\hat{h}_{\theta} (x_j) > \hat{h}_{\theta} (x_i))\cdot\Delta_{j}}{\sum_{i,j}\mathds{1}(T_i > T_j)\cdot\Delta_{j}}
\end{equation}
The CI estimates the probability of concordance $P(\hat{h}_{\theta} (x_j) > \hat{h}_{\theta} (x_i) | T_i > T_j)$ for two independent observations/predictions. It can also be interpreted as a measure of the area under a time-dependent receiver operator curve \cite{hajian2013receiver}  \cite{schmid2016CIndex},\cite{heagerty2005survival}. A value of $CI=1$ means that all observations are correctly sequenced, $CI=0.5$ means that the method applied is no better than guessing.

\subsubsection{Area Under the Receiver Operating Characteristic}
In the evaluation of the results our methods generate an important measurement value is the area under the receiver operating characteristic (AUROC)\cite{hanley1982meaning}. For a data point with $a = P(F_p)$ the probability of the data point being a false positive prediction and $b-1 = P(T_p)$ the negative probability of the data point being a true positive prediction in a dataset $D$, \cite{bradley1997use} defines the AUROC as follows:

$$\mathrm{AUROC} = \sum\limits_{i\in D}\{(1-b_i\cdot\Delta a) + \frac{1}{2}[\Delta(1-b)\cdot\Delta a]\}$$
with 
$$\Delta(1-b) = (1-b_i)-(1-b_{i-1})$$
and
$$\Delta a = a_i - a_{i-1}$$


\section{Results}
\label{sec:results_all}
An overview of all the concordance indices and AUROC values of our experiments is provided in table \ref{tab:allC_indices}.

\begin{table*}[]
    \centering
    \begin{tabularx}{\linewidth}{|*{9}{Y|}}
        \cline{2-9}
        \multicolumn{1}{c|}{} & \multicolumn{4}{| c |}{Without regularization} & \multicolumn{4}{| c |}{With regularization}\\
        \cline{2-9}
        \multicolumn{1}{c|}{} & \multicolumn{2}{| c |}{Isolated} & \multicolumn{2}{| c |}{Merged} & \multicolumn{2}{| c |}{Isolated} & \multicolumn{2}{| c |}{Merged} \\
        \cline{2-9}
        \multicolumn{1}{c|}{} & A \nolinebreak \scriptsize{(test)} & B & Naive & Standardised & A \nolinebreak \scriptsize{(test)} & B & Naive & Standardised\\
        \hline
        C-Index & 0.677 & 0.612 & 0.569 & 0.644 & 0.795 & 0.615 & 0.646 & 0.676\\
        \hline
        AUROC & 0.615 & 0.635 & 0.544 & 0.614 & 0.789 & 0.578 & 0.601 & 0.642\\
        \hline

    \end{tabularx}
    \caption{Evaluation of receiver operator curve and Harrel's C-Index with and without regularization on different testsets}
    \label{tab:allC_indices}
\end{table*}

\subsection{Results of model without regularization}
\label{sec:results_without_reg}

In our first set of experiments, we use the loss function supplied in equation (\ref{eq:loss_function}). Although we achieve an AUROC value of $61.5\%$ on the test subset of dataset A and also an AUROC value of $63.5\%$ on the separated dataset B. We observe that the domains in which our predictions $\hat{h}_{\theta} (x_i)$ lay differ significantly from dataset A to dataset B. 

\begin{figure}
    \centering
    \includegraphics[width=0.93\linewidth]{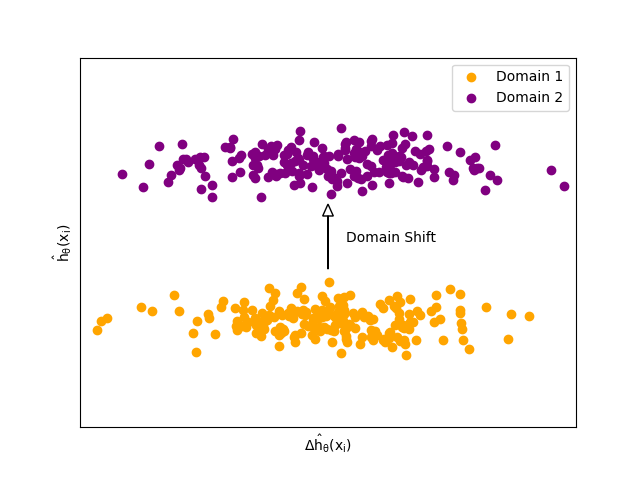}
    \caption{Domain shift in predicting on different datasets.} 
    \label{fig:domainshift}
\end{figure}

This results in a bad overall AUROC value of $54.4\%$ if one mixes these predictions naivly together as it would be a complete dataset from only one datasource before evaluating the whole set.  One potential reason is that the net learns features that are relevant for the task, but it also is sensitive to further properties  of the image like the pixel's brightness (or in general the pixel's color distribuition). While such image features do not encode meaningful medical information, they can still disturb the outcome of the network, especially if the hazards predicted on the second data set are in a different numerical range as compared with the original training data. This in particular happens by deviation from the neutral direction $\hat h_\theta(x_j)\to \hat h_\theta(x_j)+z$ which merely leads to a redefinition of the baseline hazard function $h_0(t)$ but is not sensed in the Cox loss function. So one can imagine the predicitons $\hat{h}_{\theta} (x_i)$ shifted away from the total diagonal like schematically shown in figure \ref{fig:domainshift}. 

In our case, dataset B has a slightly other color scheme (mainly because of the use of another scanner - see section \ref{sec:DataDescription}) and so our predictions evaluating the model trained on dataset A on this dataset  leads to a significant drop in performance.

A first approach to reduce the sensitivity towards different hazards is to re-center the predicted hazards. We thus standardise (we subtract the mean and divide the result by the standard deviation) the predictions of each dataset and merge them afterwards. We display the resulting improvement in the rightest column of each approach in table \ref{tab:allC_indices}.

The downside of this approach is that you need to have already a certain set of of images for each source of images/scanner that is used to predict survival probabilities. 

\subsection{Results of model with regularization}
\label{sec:results_with_reg}

To interdict shifting of hazard functions alltogether, we add a L2-regularization-term, see equation (\ref{eq:lossWithReg}), to shift all predictions $\hat{h}_{\theta} (x_i)$ in the same domain range. This does not only lead to a higher AUROC value in evaluation on the mixed data set ($60.1\%$ instead of $54.4\%$), but we can improve even more if combining the regularization with the above-mentioned normalization process ($64.2\%$ instead of $61.4\%$)

Besides, we also achieved a significantly higher ROC value on the testdata of dataset A throughout this regularization technique. Unfortunately the overall generalizability on the isolated dataset B suffered ($57.8\%$ instead of $63.5\%$) from that approach.



\section{Discussion and Outlook}
\label{sec:discussion}
We have shown that it is possible to make survival predictions based on simplified image information using Cox's Proportional Hazard Models on neural networks and that our domain adaptation techniques succeed in merging the predictions into one range.
Compared to \cite{li2022application} where the authors achieved an AUROC value of 76.9\% our approach with regularization slightly outperforms it with an AUROC value of 79.5\%.

In future work, we will have a more detailed evaluation and we have a closer look on levels of significance and correlations to clinical variables like the Breslow depth or genetic information and demographic variables such as age or gender to evaluate and improve the clinical utility of our predictions.



\printbibliography
\begin{table*}[]
\centering
 \begin{tabular}{|c|c|c c c|c|} 
 \hline
 \multicolumn{2}{|c|}{Characteristic} & \multicolumn{3}{c|}{Dataset A } & Dataset B \\
  \multicolumn{2}{|c|}{} & Training & Validation & Test & Test \\ 
 \hline
 \multirow{2}{*}{N} & Scans & 313 & 36 & 38 & 242 \\
  & Patients & 104 & 36 & 36 & 242 \\
  \hline
 Age & Median & 60 & 61 & 57.5 & 70 \\ 
 \hline
 \multirow{2}{*}{Sex} & Male & 55 & 19 & 19 & 132 \\
  & Female & 49 & 17 & 17 & 110 \\
  \hline
 \multirow{10}{*}{AJCC Stage} &  IA & 18 (17\%) & 6 (17\%) & 8 (22\%) & 0 (0\%) \\
  & IB       & 32 (31\%) & 10 (28\%) & 9 (25\%) & 0 (0\%)\\
  & IIA      & 11 (11\%) & 5 (14\%)  & 3 (8\%) & 116 (48\%)\\
  & IIB      & 5 (5\%)   & 4 (11\%)  & 5  (14\%) & 78 (32\%)\\
  & IIC      & 6 (6\%)   & 0 (0\%)   & 0 (0\%) & 48 (20\%)\\
  & IIIA/B/C & 2 (2\%)   & 2 (6\%)   & 2 (6\%) & 0 (0\%)\\
  & IIIA     & 6 (6\%)   & 3 (8\%)   & 2 (6\%) & 0 (0\%)\\
  & IIIB     & 7 (7\%)   & 2 (6\%)   & 2 (6\%) & 0 (0\%)\\
  & IIIC     & 13 (13\%) & 4 (11\%)  & 5 (14\%) & 0 (0\%)\\
  & IIID     & 4 (4\%)   & 0 (0\%)   & 0 (0\%) & 0 (0\%)\\ 
 \hline
 \multirow{2}{*}{Breslow thickness} & Mean [mm] & 2.8 & 1.67 & 2.0 & 3.72\\
  & Median [mm] & 1.7 & 1.55 & 1.2 & 3\\
  \hline
 \multirow{8}{*}{Primary Tumor} & T1a & 9 (9\%) & 4 (11\%) & 2 (6\%) & 0 (0\%) \\
 & T1b & 12 (12\%) & 4 (11\%) & 8 (22\%) & 0 (0\%) \\
 & T2a & 37 (36\%) & 13 (36\%) & 9 (25\%) & 0 (0\%) \\
 & T2b & 5 (5\%) & 7 (19\%) & 4 (11\%) & 33 (14\%) \\
 & T3a & 14 (13\%) & 1 (3\%) & 4 (11\%) & 83 (34\%) \\
 & T3b & 9 (9\%) & 6 (17\%) & 6 (17\%) & 60 (25\%) \\
 & T4a & 3 (3\%) & 1 (3\%) & 1 (3\%) & 18 (7\%) \\
 & T4b & 15 (14\%) & 0 (0\%) & 2 (6\%) & 48 (20\%) \\
 \hline
 \multirow{7}{*}{Node Status} & N0 & 67 (64\%) & 25 (69\%) & 25 (69\%) & 242 (100\%)\\
 & N1a & 1 (1\%) & 1 (3\%) & 0 (0\%) & 0 (0\%)\\
 & N1b & 1 (1\%) & 0 (0\%) & 0 (0\%) & 0 (0\%)\\
 & N2b & 0 (0\%) & 1 (3\%) & 0 (0\%) & 0 (0\%)\\ 
 & N3 & 5 (5\%) & 1 (3\%) & 0 (0\%) & 0 (0\%)\\
 & N3c & 3 (3\%) & 0 (0\%) & 1 (3\%) & 0 (0\%)\\
 & unknown & 27 (26\%) & 9 (25\%) & 10 (28\%) & 0 (0\%)\\
 \hline
 \multirow{4}{*}{Sentinel Lymph Node Status} & Negative & 68 (65\%) & 25 (69\%) & 25 (69\%) & 194 (80\%)\\
  & Positive & 30 (29\%) & 11 (31\%) & 10 (28\%) & 0 (0\%)\\
  & Unknown & 6 (6\%) & 0 (0\%) & 1 (3\%) & 3 (1\%)\\
  & Not performed & 0 (0\%) & 0 (0\%) & 0 (0\%) & 45 (19\%)\\
 \hline
  \multirow{2}{*}{Ulceration} & Absent & 68 (65\%) & 22 (61\%) & 23 (64\%) & 101 (42\%)\\
  & Present & 36 (35\%) & 14 (39\%) & 13 (36\%) & 141 (58\%)\\
 \hline
 \end{tabular}
 \caption{\textbf{Appendix A:} Additional features of our datasets.}
 \label{tab:clinical_table_detailed}
\end{table*}

\end{document}